\documentclass[8.5pt,twoside,twocolumn]{article}
\usepackage[super,sort&compress,comma]{natbib} 
\usepackage{mhchem}
\usepackage{times,mathptmx}
\usepackage{sectsty}
\usepackage{balance}
\usepackage{color,bm,epsfig,graphics,amssymb,amsmath,subeqnarray,setspace,graphicx,amsthm,epstopdf,subfigure,color} 

\usepackage{bm,epsfig,graphics,amssymb,amsmath,subeqnarray,setspace,graphicx,amsthm,epstopdf,subfigure,color}
\usepackage{epstopdf, epsf,psfrag, graphicx,color}
\usepackage{widetext}

\usepackage{graphicx} 
\usepackage{lastpage}
\usepackage[format=plain,justification=raggedright,singlelinecheck=false,font=small,labelfont=bf,labelsep=space]{caption} 
\usepackage{fancyhdr}
\usepackage{bm,epsfig,graphics,amssymb,amsmath,subeqnarray,setspace,graphicx,amsthm,epstopdf,subfigure,color}
\usepackage{epstopdf, epsf,psfrag, graphicx,color}
\usepackage{float}
\usepackage{multicol}

\floatstyle{plain}
\newfloat{twocolequfloat}{b}{zzz}
\floatname{twocolequfloat}{Equation}

\definecolor{DarkBlue}{rgb}{0,0,.8}

\newcommand\ski{}
\newcommand\tom{}
\newcommand\TP{}

\def\Er{\mathrm{Er}}
\def\b{\mathbf}
\def\e{\varepsilon}
\def\n{\mathbf{n}}

\renewcommand{\Re}{\operatorname{\mathrm{Re}}}

\begin{document}

%
%

\makeatletter 
\def\subsubsection{\@startsection{subsubsection}{3}{10pt}{-1.25ex plus -1ex minus -.1ex}{0ex plus 0ex}{\normalsize\bf}} 
\def\paragraph{\@startsection{paragraph}{4}{10pt}{-1.25ex plus -1ex minus -.1ex}{0ex plus 0ex}{\normalsize\textit}} 
\renewcommand\@biblabel[1]{#1}            
\renewcommand\@makefntext[1]%
{\noindent\makebox[0pt][r]{\@thefnmark\,}#1}
\makeatother 
\renewcommand{\figurename}{\small{Fig.}~}
\sectionfont{\large}
\subsectionfont{\normalsize}

\fancyfoot{}
\fancyfoot[LO,RE]{\vspace{-7pt}\includegraphics[height=9pt]{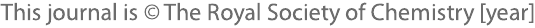}}
\fancyfoot[CO]{\vspace{-7.2pt}\hspace{12.2cm}\includegraphics{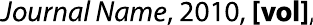}}
\fancyfoot[CE]{\vspace{-7.5pt}\hspace{-13.5cm}\includegraphics{RF}}
\fancyfoot[RO]{\footnotesize{\sffamily{1--\pageref{LastPage} ~\textbar  \hspace{2pt}\thepage}}}
\fancyfoot[LE]{\footnotesize{\sffamily{\thepage~\textbar\hspace{3.45cm} 1--\pageref{LastPage}}}}
\fancyhead{}
\renewcommand{\headrulewidth}{1pt} 
\renewcommand{\footrulewidth}{1pt}
\setlength{\arrayrulewidth}{1pt}
\setlength{\columnsep}{6.5mm}
\setlength\bibsep{1pt}

\twocolumn[
  \begin{@twocolumnfalse}
\noindent\LARGE{\textbf{Microscale locomotion in a nematic liquid crystal}}
\vspace{0.6cm}

\noindent\large{\textbf{Madison S. Krieger\textit{$^{a}$}, Saverio E. Spagnolie\textit{$^{b}$}, and Thomas Powers\textit{$^{a,c}$}}}\vspace{0.5cm}

\noindent\textit{\small{\textbf{Received Xth XXXXXXXXXX 20XX, Accepted Xth XXXXXXXXX 20XX\newline
First published on the web Xth XXXXXXXXXX 20XX}}}

\noindent \textbf{\small{DOI: 10.1039/b000000x}}
\vspace{0.6cm}

\noindent \normalsize{Microorganisms often encounter anisotropy, for example in mucus and biofilms. We study how anisotropy and elasticity of the ambient fluid affects the speed of a swimming microorganism with a prescribed stroke. Motivated by recent experiments on swimming bacteria in anisotropic environments, we extend a classical model for swimming microorganisms, the Taylor swimming sheet, actuated by 
\tom{small-amplitude} traveling waves in a three-dimensional nematic liquid crystal without twist. We calculate the swimming speed and entrained volumetric flux as a function of the swimmer's stroke properties as well as the elastic and rheological properties of the liquid crystal. \ski{{These results are then compared} to previous {results} on an analogous swimmer in a hexatic liquid crystal, {indicating} large differences in the cases of small Ericksen number and in a nematic fluid {when the} tumbling parameter {is} near the transition to a {shear}-aligning nematic.} We also propose a novel method of \tom{swimming in a} 
nematic fluid by passing a traveling wave of director oscillation along a rigid wall.}
\vspace{0.5cm}
 \end{@twocolumnfalse}
]

\footnotetext{\textit{$^{a}$~School of Engineering, Brown University, Providence, RI 02912 USA. Email: madison\_krieger@brown.edu}}
\footnotetext{\textit{$^{b}$~Department of Mathematics, University of Wisconsin-Madison, Madison, WI 53706, USA. Email: spagnolie@math.wisc.edu}}
\footnotetext{\textit{$^{c}$~Department of Physics, Brown University, Providence, RI 02912, USA. Email: thomas\_powers@brown.edu}}

\section{Introduction}
The nature of the fluid through which a microorganism swims has a profound effect on strategies for locomotion. 
At the small scale of a bacterial cell, inertia is unimportant and locomotion is constrained by the physics of low-Reynolds-number\footnote{The Reynolds number Re for the flow of a fluid with viscosity $\mu$, density $\rho$, characteristic flow length $L$, and characteristic flow velocity $v$ is $\mathrm{Re}=\rho v L/\mu$.} flows~\cite{taylor1951,purcell1977,LaugaPowers2009}. In a Newtonian liquid such as water, low-Reynolds number locomotion is characterized by two distinctive properties: a vanishingly small timescale for the diffusion of velocity, and drag anisotropy, which is a difference between the viscous drag per unit length on a thin filament translating along its long axis and
transverse to its long axis~\cite{LaugaPowers2009}. In resistive force theory, drag anisotropy is required for locomotion~\cite{hancock53,becker03,PakLauga2011}. 

\TP{In complex fluids such as polymer solutions and gels, the elasticity of the polymers introduces a new timescale, the elastic relaxation timescale, which is much longer than the timescale for the diffusion of velocity~\cite{doi_edwards1986}.}  When the fluid has an elastic response to deformation, swimming speeds can increase or decrease depending on the body geometry and the elastic relaxation timescale~\cite{Lauga2007,FuPowersWolgemuth2007,TeranFauciShelley2010,ShenArratia2011, LiuPowersBreuer2011,dasgupta_etal2013,slp13,tg14,tg14b}, and the so-called scallop theorem does not apply \cite{NormandLauga2008,FuPowersWolgemuth2008Scallop}. 
Swimmers can move faster in gels and networks of obstacles than in a Newtonian liquid~\cite{Leshansky2009,FuShenoyPowers2010,manlauga15}. 
When the flagellum size is similar to the size of the polymers, local shear-thinning may be the primary cause of swimming speed variations in such fluids~\cite{RodrigoLauga2013,Montenegro-Jetal2013,martinez2014flagellated,lka14}.

\begin{figure*}[th]
\includegraphics[width=\textwidth]{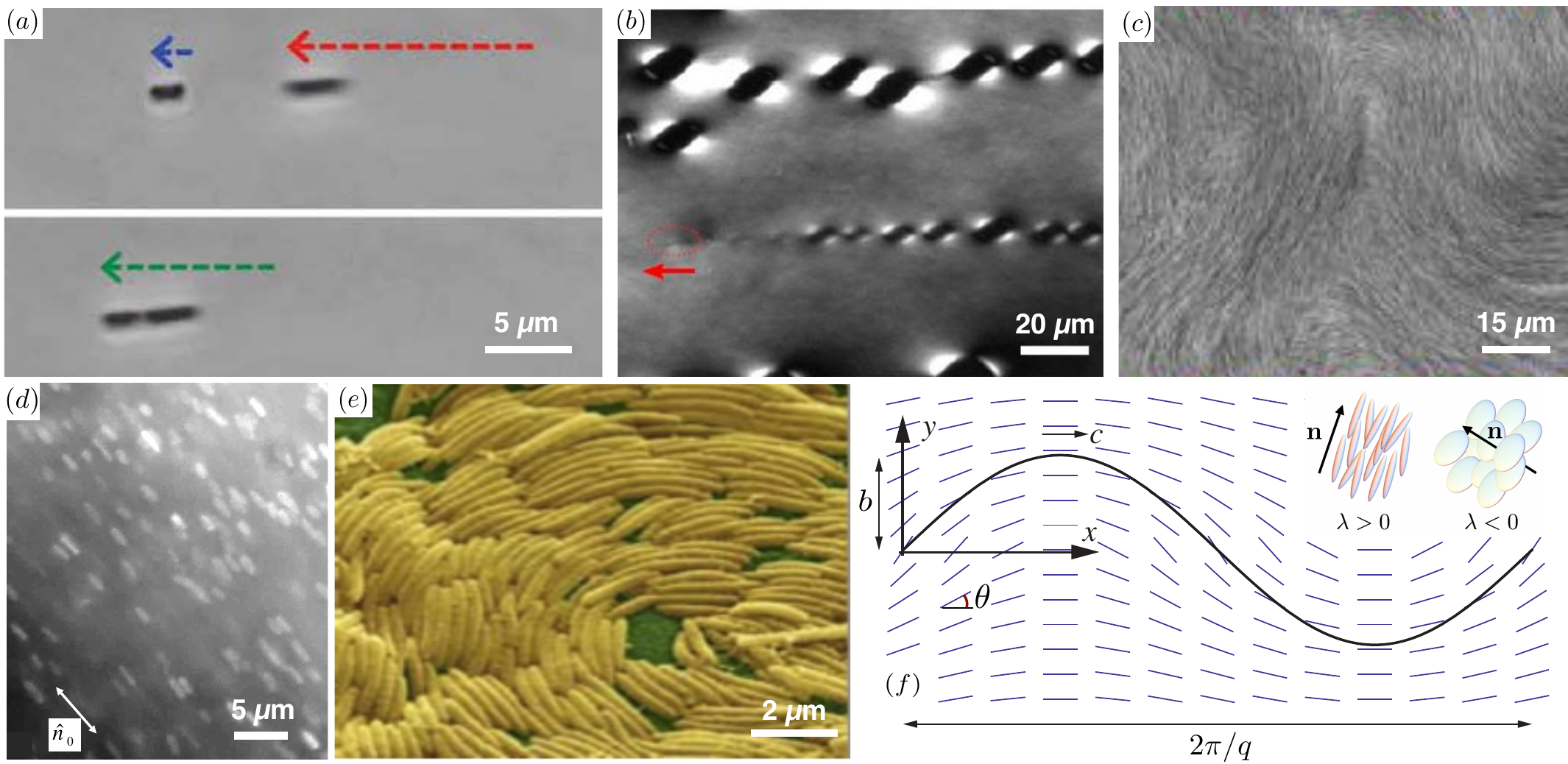}
\caption{(Color online) (a) Dynamic assembly of bacteria in disodium \TP{cromoglycate}~\cite{mttwa14}. 
(b) Melting of the liquid crystal medium behind a flagellated swimmer~\cite{zsla14}. (c) Disclinated texture observed as a collection of bacteria locomote in a nematic liquid crystal~\cite{zsla14}. (d) Aligned \TP{\textit{Pseudomonas aeruginosa}} cells in the 
\TP{liquid crystal} matrix of concentrated DNA~\cite{Smalyukh_etal2008}. (e) Aligned swimmers in a bacterial flock. Figure from Gregory Velicer (Indiana University Bloomington)
and Juergen Bergen (Max-Planck Institute for Developmental Biology). (f) Sketch of a swimming sheet (not to scale) immersed in a nematic liquid crystal with director field $\n(x,y,t)$. 
The propagating wave has wavelength $2\pi/q$, small amplitude $b\ll2\pi/q$, and wave speed $c=\omega/q$. The director field $\n$ makes an angle $\theta$ with the $x$ axis.}
\label{setup}
\end{figure*}

\TP{Like polymer solutions and gels, liquid crystals have an elastic relaxation time scale, but they also alter the drag anisotropy required for propulsion since the fluid itself exhibits anisotropy. For example the nematic liquid crystal phase consists of rod-like molecules which spontaneously align in the absence of an external field. }
The consequences of 
\TP{molecular} anisotropy on the locomotion of microorganisms have recently been explored experimentally.~{\it Proteus mirabilis} cells were found to align with the 
\TP{nematic} director field and form multi-cellular assemblies~\cite{mttwa14,mtwa14,kumar2013LCswim} (Fig. 1a). When swimming near 
\TP{nematic} droplets, surface topological defects were shown to play an important role in bacterial escape from the 
\TP{liquid crystal} interface~\cite{mtwa14}.  Collective dynamic effects and director-guided motion was also observed in {\it Bacillus subtilis} at low bacterial volume fraction, and a local melting of the 
\TP{liquid crystal} caused by the bacteria was found~\cite{zsla14} (Fig.~1b,c). Potential applications include the delivery of small cargo using the direction of molecular orientation~\cite{szla15}. Understanding these results may be relevant in understanding locomotion in biofilms~\cite{Smalyukh_etal2008} (Fig.~1d), and is complementary to recent work on active nematics, or soft active matter, in which dense suspensions of microorganisms themselves can exhibit \TP{nematic}-like ordering~\cite{TonerTuRamaswamy2005,mjrlprs13,ss15} (Fig. 1e).

A classical mathematical model of swimming microorganisms is Taylor's swimming sheet\cite{taylor1951}, in which either transverse or longitudinal waves of small amplitude propagate along an immersed sheet of infinite extent. Extensions of this model have been used to study other important phenomena such as hydrodynamic synchronization \cite{Fauci90,el09,ElfringPakLauga2010,el11b}, interactions with other immersed structures \cite{cfs13,dp13} and geometric optimization \cite{mjl14}. Other variations on this asymptotic model have been used to study locomotion in a wide variety of complex fluids by numerous authors~\cite{elfring2015theory}. Locomotion in liquid crystals, however, has not yet seen much theoretical treatment. In previous works~\cite{ksp14,kdp15}, we studied a one-dimensional version of Taylor's swimming sheet in a two-dimensional hexatic LC film. Departure from isotropic behavior in that model is greatest for large rotational viscosity and strong anchoring boundary conditions, and the swimming direction depends on fluid properties. Further unusual properties for Taylor's swimming sheet were observed, such as the presence of a net volumetric flux. Because the nematic phase is more commonly observed than the hexatic, the present study is intended \tom{to explore new features that arise with nematic order, and} also to determine \tom{when, if ever, the hexatic model can be used to accurately describe swimming in a nematic liquid crystal.}

In this article we extend the Taylor swimming sheet model to the study of force- and torque-free undulatory locomotion in a three-dimensional nematic liquid crystal, with tangential anchoring of arbitrary strength on the surface of the swimmer.
\TP{We assume the director lies in the $xy$-plane  and does not twist (Fig.~\ref{setup}f).}
\ski{Alternatively the problem could be considered as filament motion in a two-dimensional nematic fluid.}
By performing an asymptotic calculation to second-order in the wave amplitude, assumed small compared to the wavelength, we examine how fluid anisotropy and relaxation affects swimming speed.
We show how the swimming velocity depends on numerous physical parameters, such as the rotational viscosity $\gamma$, anisotropic viscosities $\mu_i$, the Frank \TP{elastic} constants $K_i$, the tumbling parameter $\lambda$, and the Ericksen number $\Er$, which measures the relative viscous and elastic forces in the fluid.   \color{black}The rate of fluid transport \tom{induced by swimming} is also investigated; 
unlike in a Newtonian fluid, \tom{the induced fluid flux} can 
\tom{be either along} 
or against the motion of the swimmer. 


The paper is organized as follows: In \S\ref{sec:VEstress} we describe the stresses that arise in a continuum treatment of a nematic liquid crystal near equilibrium. In \S\ref{sec:goveqn} we use these stresses to derive a set of coupled equations for the flow field and local nematic orientation. Following Taylor~\cite{taylor1951}, we nondimensionalize and expand these equations perturbatively to first- and second-order in 
wave amplitude and derive an integral relation for the swimming speed and volume flux in \S\ref{sec:leading} and \S\ref{sec:secondary}. The dependence of the swimming speed and flux on Ericksen number, rotational viscosity, and tumbling parameter is described in \S\ref{sec:results}.
In \S\ref{sec:noshape}, we show that a propagating wave of director oscillation can result in fluid pumping and locomotion of a passive flat surface. \ski{To determine the regimes in which the results for swimming speed and flux are comparable in nematic and hexatic fluids, and where they differ, we plot these quantities side-by-side and discuss the results in the Discussion, \S\ref{sec:discuss}.}  


\section{Theory}
\label{sec:theory}

\subsection{Viscous and elastic stresses} \label{sec:VEstress} In a continuum treatment of a nematic liquid crystal, a local average of molecular orientations is described by the director field $\n$. The fluid's viscous stress response to deformation is approximated by incorporating terms linear in the strain rate that preserve $\n\rightarrow-\n$ symmetry. In an incompressible nematic, 
the deviatoric viscous stress~\cite{larson1999,landau_lifshitz_elas} is
\begin{equation}
\bm{\sigma}^\mathrm{d}=2\mu\mathsf{E}
+2\mu_1^*\n\n\left( \n\cdot\mathsf{E}\cdot\n\right)
+\mu_2^*\left(\n\mathsf{E}\cdot\n+\n\cdot\mathsf{E}\n
\right), \label{viscstress}
\end{equation}
with $\mathsf{E}= \left[\bm{\nabla}\mathbf{v}+(\bm{\nabla}\mathbf{v})^\mathrm{T}\right]/2$ the symmetric rate-of-strain tensor, and $\mathbf{v}$ the velocity field. The shear viscosity of an isotropic phase is $\mu$, and $\mu_1^*$ and  $\mu_2^*$ are viscosities arising from the anisotropy. The coefficients $\mu_1^*$ and $\mu_2^*$ can be negative, but the physical requirement that the power dissipation be positive yields bounds of $\mu>0$, $\mu_2^*>-2\mu$, and $\mu_1^*+\mu_2^*>-3\mu/2$. \ski{While both the nematic phase and the hexatic phase we studied previously~\cite{ksp14} are anisotropic, the hexatic phase has $\mu_1^*=\mu_2^*=0$ and thus} has an isotropic viscous stress tensor, in contrast with the nematic. 

Meanwhile, the elastic free energy for a nematic liquid crystal is
\begin{equation}
 \mathcal{F}=\frac{K_1}{2}\left(\bm{\nabla}\cdot\n\right)^2+\frac{K_2}{2}\left(\n\cdot\bm{\nabla}\times\n\right)^2+
 \frac{K_3}{2}\left[\n\times\left(\bm{\nabla}\times\n\right)\right]^2,
 \label{FrankFE}
 \end{equation}
where $K_1$ is the splay elastic constant, $K_2$ is the twist elastic constant, and $K_3$ is the bend elastic constant~\cite{larson1999,landau_lifshitz_elas}. The total free energy in the fluid (per unit length) is $F_\mathrm{el}=\int\mathcal{F}\mathrm{d}x\mathrm{d}y$. 
As mentioned earlier, for simplicity we do not consider twist, and thus we disregard $K_2$. Thus, the angle field $\theta(x,y,t)$ 
completely determines the nematic configuration (Fig.~1f). \tom{Comparing again with our previous study~\cite{ksp14},  there is only one Frank constant when the two-fold symmetry of the nematic is enlarged to the six-fold symmetry of a hexatic.}

Equilibrium configurations of the director field are found by minimizing $\mathcal{F}$ subject to $|\n|=1$. This procedure leads to $\mathbf{h}=0$, where $\mathbf{h}$ is the transverse part of the molecular field $\mathbf{H}=-\delta F_\mathrm{el}/\delta{\n}$; $\mathbf{h}=\mathbf{H}-\n\n\cdot\mathbf{H}$.
Near equilibrium, the fluid stress corresponding to the elastic free energy $\mathcal{F}$ is then~\cite{landau_lifshitz_elas,deGennesProst}
 \begin{equation}
 \sigma^\mathrm{r}_{ik}=-\Pi_{kl}\partial_i n_l-\frac{\lambda}{2}\left(n_i h_k + n_k h_i\right)
 +\frac{1}{2}\left(n_i h_k - n_k h_i\right),\label{elastress}
 \end{equation}
where $\Pi_{ki}=\partial F_\mathrm{el}/{\partial(\partial_k n_i)}$. \ski{In equilibrium, the condition for the balance of director torques $\mathbf{h}=\mathbf{0}$ implies the balance of elastic forces, $-\partial_i p_\mathrm{eq}+\partial_j\sigma_{ij}^\mathrm{r}=0$, provided the pressure is given by $p_\mathrm{eq}=-\mathcal{F}$~\cite{deGennesProst}. The ``tumbling parameter" $\lambda$ is not a dissipative coefficient, but is related to the degree of order and the type of nematic, with calamitic phases (composed of rod-like molecules) tending to have $\lambda > 0$, and discotic phases (composed of disk-like molecules) tending to have $\lambda < 0$. The value of this parameter further classifies nematic fluids as either ``tumbling'' ($\lambda < 1$) or ``{shear}-aligning'' ($\lambda \geq 1$). In a simple shear flow, tumbling nematics continuously rotate whereas {shear}-aligning nematics tend to align themselves at a certain fixed angle relative to the principal direction of shear. In DSCG, a lyotropic chromonic liquid crystal commonly used in experiments on swimming microorganisms in liquid crystals, the tumbling parameter $\lambda$ is a function of temperature and has a range $\lambda = 0.6-0.9$~\cite{yaoPhD}. For comparison, the hexatic phase has $\lambda = 0$ and therefore lacks any of these distinctions.}

\tom{Alternatively, we could have formulated the stresses using the Ericksen-Leslie approach; the connection between the Ericksen-Leslie approach and that used here is discussed in the references~\cite{deGennesProst,KlemanLavrentovich2003}.}

\subsection{Governing equations}\label{sec:goveqn} The swimming body is modeled as an infinite sheet undergoing a prescribed transverse 
sinusoidal undulation of the form 
$Y^* = (\varepsilon/q)\sin(qx-\omega t)$,
measured in the frame moving with the swimmer. Here $\varepsilon$ 
is the dimensionless amplitude  for the swimmer.
We focus on transverse waves in the body of this article, but we briefly treat longitudinal waves in the appendix.

At zero Reynolds number, conservation of mass of an incompressible fluid results in a divergence-free velocity field, $\nabla \cdot \b{v}=0$, and conservation of momentum is expressed as force balance,
\begin{equation}
-\partial_ip+\partial_j\left(\sigma^{\mathrm d}_{ij}+\sigma^\mathrm{r}_{ij}\right)=0.
\label{fbal}
\end{equation}
Torque balance is expressed by~\cite{larson1999,landau_lifshitz_elas}
\begin{eqnarray}
\partial_t n_i+\left(\mathbf{v}\cdot\bm{\nabla}\right)n_i
-\frac{1}{2}\left[\left(\bm{\nabla}\times\mathbf{v}\right)\times\n\right]_i \nonumber  \\
={\lambda}\left(\delta_{ij}-n_in_j\right)E_{jk} n_k+h_i/\gamma ^*,\label{dndt}
\end{eqnarray}
where $\gamma ^*$ is a rotational or twist viscosity\footnote{\ski{For  comparison with the much simpler hexatic phase, Appendix~\ref{hexaticappx} includes the governing equations for a hexatic liquid crystal~\cite{ksp14}.}}. In DSCG, $\gamma^*/\mu$ ranges from $\approx 5$ to $\approx 50$~\cite{DSCGprop}. The viscous torque arising from the rotation of the director relative to the local fluid rotation balances with viscous torque arising through $\mathsf{E}$ and elastic torque through $-\mathbf{h}$. We work in the rest frame of the swimmer.

The no-slip velocity boundary condition is applied on the swimmer surface, and as $y\rightarrow\infty$ the flow has uniform velocity $\mathbf{v}=U^* \hat{\mathbf{x}}$ where $-U^*$ is the swimming speed. Meanwhile, the director field has a preferential angle at the boundary due to anchoring conditions. We will study the case of tangential anchoring at the swimmer surface, with anchoring strength $W$\cite{deGennesProst}. Since we expand in powers of the amplitude, we may write this condition to second order in the angle field: 
\begin{equation}
-K_1\partial_y\theta+W(\theta- \partial_xY^*)=0,\label{tangbc}
\end{equation}
where $y=Y^*(x,t)$ describes the swimmer shape, and (\ref{tangbc}) is evaluated at $y=Y^*$. It is convenient to define the dimensionless anchoring strength $w=W/(q K_1)$. 

Henceforth we treat $x$, $y$, and $t$ as dimensionless variables by measuring length in units of $q^{-1}$ and time in units of $\omega^{-1}$.
The dimensionless viscosities are defined by $\mu_1 = \mu_1 ^*/\mu$,  $\mu_2 = \mu_2 ^*/\mu$ and $\gamma = \gamma ^*/\mu$. It is also convenient to introduce the wave speed $c=\omega/q$, which is one in the natural units. 
The ratio of Frank constants is denoted by $K_r=K_1 / K_3 $, and we define 
$U = U^* / c$, and $Q = Q^*/(\omega \e ^2/q^2)$ for the 
volumetric flux. The undulating shape of the swimmer takes the nondimensional form 
\begin{equation}
(X,Y) = (0, \varepsilon)\sin(x- t). \label{yform}
\end{equation}
The elastic response of the fluid to deformation introduces a length-scale-dependent relaxation time, $\tau=\mu/(K_3 q^2)$. For small-molecule liquid crystals, typical values are $\mu\approx 10^{-2}$\,$\mathrm{Pa\,s}$ and $K_3\approx10^{-11}$\,N. On the length scale of bacterial flagellar undulations for which $q\approx1\,\mu$m$^{-1}$, the relaxation time is $\tau\approx1$\,ms. Comparing the typical viscous stress \eqref{viscstress} with the typical elastic stress \eqref{elastress}, we find the Ericksen number~\cite{larson1999}, written  $\Er=\tau\omega$. Note that unlike the Reynolds number, which is always small for swimming microorganisms, the Ericksen number for a swimming microorganism may be small or large. The beat frequencies and wavenumbers of undulating cilia and flagella vary widely~\cite{Lauga2007,smith2009bend}, and for experiments  on bacteria in liquid crystals the Ericksen number can be small~\cite{kumar2013LCswim,mtwa14}, $\Er\approx10^{-1}$, or large~\cite{zsla14},  $\Er\approx10^1$.

\subsection{Leading order fluid flow}\label{sec:leading}

Following Taylor\cite{taylor1951}, 
we pursue a regular perturbation expansion in the wave amplitude $\e$. The stream function $\psi$ is defined by 
$\b{v} = \nabla\times(\psi\hat{z})$;  
this form ensures $\nabla \cdot \b{v}=0$. The stream function $\psi$ and the angle field $\theta$ are expanded in powers of $\e$ as $\psi=\e \psi^{(1)}+\e^2\psi^{(2)}+O(\e^3)$ and $\theta=\e \theta^{(1)}+\e^2\theta^{(2)}+O(\e^3)$. Force and torque balance from (\ref{fbal}) and (\ref{dndt}) at $O(\e)$ are given by 

\begin{widetext}
\begin{eqnarray}
\nabla^4\psi^{(1)}+\frac{4\mu_1}{2  + \mu _2}\partial_x^2\partial_y^2\psi^{(1)}
+\frac{1}{\mathrm{(2+\mu_2)\Er}}\left\{(1+\lambda)\partial_x^4\theta^{(1)}+\left[K_r(1+\lambda)+1-\lambda\right]\partial_x^2\partial_y^2\theta^{(1)}+K_r(1-\lambda)\partial_y^4\theta^{(1)}\right\}&=&0,
\label{stokes1stream}\\
\partial_t\theta^{(1)}+\frac{1+\lambda}{2}\partial_x^2\psi^{(1)}
+\frac{1-\lambda}{2}\partial_y^2\psi^{(1)}-\frac{1}{\Er \, \gamma}\left(\partial_x^2\theta^{(1)}+K_r\partial_y^2\theta^{(1)}\right)&=&0.\label{thetat1stream}
\end{eqnarray}
\end{widetext}
\noindent 
 Equations (\ref{stokes1stream}) and (\ref{thetat1stream}) are solved by $\psi^{(1)}=\Re[\tilde{\psi}^{(1)}]$ and $\theta^{(1)}=\Re[\tilde{\theta}^{(1)}]$, where
\begin{eqnarray}
\tilde{\psi} ^{(1)} &=& \sum_{j=1}^3 c_j e^{r_j y  + \color{black} i(x-t)}, \label{rootansatz1}\\ 
\tilde{\theta}^{(1)} &=&\sum_{j=1}^3 d_j e^{r_j y   +  \color{black} i(x-t)}.\label{rootansatz2} \end{eqnarray}
Insertion of \eqref{rootansatz1} {and \eqref{rootansatz2}} into (\ref{stokes1stream}) and (\ref{thetat1stream}) results in a cubic equation for $m=r_j^2$, 
\begin{eqnarray}
0&=& A_0 + m A_1 + m^2 A_2 + m^3 A_3, \nonumber \\
A_0 &=&-2(2+ \mu_2) + \gamma[-(1+\lambda)^2 + 2 \mathrm{i}\Er(2  + \mu_2)],  \nonumber \\
A_1 &=& m\{K_r [\gamma(1+\lambda)^2 + 2(2+ \mu_2)] + 2(4 + 4 \mu_1 + 2 \mu_2) \nonumber \\
&+& \gamma[1-\lambda ^2 - 2 \mathrm{i}\Er(2  + 2 \mu_1 + \mu_2)]\}, \nonumber \\
A_2&=& A_0 + 4 \gamma \lambda + 2 K_r [ \gamma (-1+\lambda^2) - 4  - 4 \mu_1 - 2 \mu_2], \nonumber \\
A_3 &=& K_r [\gamma (-1 + \lambda)^2 + 4  + 2 \mu_2].
 \label{rooteqn}
\label{fullrteqn}
\end{eqnarray}
The velocity field remains finite as $y\to \infty$ if the roots $r_j$ are taken with negative real part. The relation between the coefficients $c_j$ and $d_j$ 
follows from (\ref{stokes1stream}) and (\ref{thetat1stream}):
\begin{equation}
d_j=c_j \frac{\Er \, \gamma[1+\lambda - (1-\lambda)r_j ^2]}{2(1- K_r r_j ^2 -\mathrm{i} \, \Er \, \gamma)},\label{alg1cond} 
\end{equation}
and the coefficients $c_i$ are determined by the boundary conditions at first order in amplitude
\begin{eqnarray}
\partial _y \tilde{\psi} ^{(1)} |_{y=0}& =& 0, \\
-\partial _x \tilde{\psi}^{(1)} |_{y=0} &= &- \varepsilon e^{i(x-t)}, \\
 -\partial _y \tilde{\theta}^{(1)} + w \tilde{\theta}^{(1)} |_{y=0}& =& w \varepsilon e^{i(x-t)} \label{firstanchcond}.
\end{eqnarray}

\begin{figure*}[t]
\includegraphics[width=\textwidth]{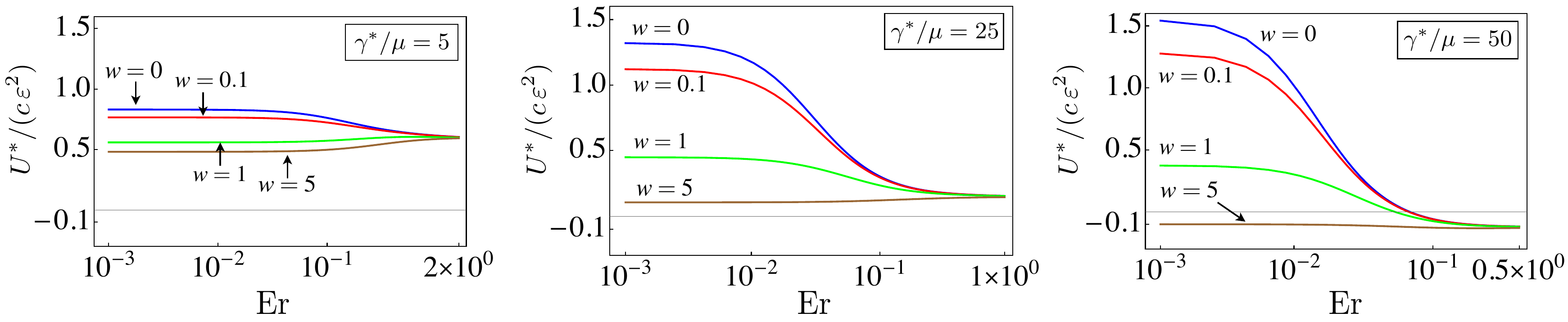}
\caption{(Color online) Dimensionless swimming speed $U^*/(c \varepsilon^2)$  vs $\Er = \tau \omega$ for a swimmer in a nematic liquid crystal with  $\mu_1^*=\mu$, $\mu_2^*=\mu$,
 $K_r = 1.2$, and $\lambda = 0.6$, with anchoring strengths $w=0$ (blue), $w=0.1$ (red), $w=1$ (green), and $w=5$ (brown). The rotational viscosity $\gamma^*/\mu$ is $5$ (left), $25$ (center), and $50$ (right). The horizontal axis is rescaled in each plot for enhanced resolution.}
\label{gamtryp}
\end{figure*}

\subsection{Second-order problem}\label{sec:secondary}

The equations at second order in $\e$ have many terms and are unwieldy. However, they are simplified by 
averaging over the spatial period. Since the forcing is a traveling sinusoidal wave depending on space and time through the combination $x-t$, {averaging over $x$ causes derivatives with respect to $x$ and with respect to $t$ to vanish.}
Denoting the spatial average by $\langle \cdot \rangle$ , we find
\begin{eqnarray}
\frac{(1-\lambda)K_r}{(2+\mu_2)\Er}\langle\partial_y^3\theta^{(2)}\rangle+\langle\partial_y^2v_x^{(2)}\rangle&=&f,\label{2o1}\\
\frac{K_r}{\gamma \, \Er}\langle\partial_y^2\theta^{(2)}\rangle-\frac{1}{2}(1-\lambda)\langle\partial_y v_x^{(2)}\rangle&=&g, \label{2o2}
\end{eqnarray}
where $f$ and $g$ are given by 
\begin{eqnarray}
f=\frac{k_1}{\Er}\langle\partial_x\theta^{(1)}\partial _y^{2}\theta^{(1)}\rangle
+\frac{4\mu_1}{2+\mu_2}\langle\bm{\nabla}\theta^{(1)}\cdot\partial_y\mathbf{v}^{(1)}\rangle \label{ff} \\
g=\langle\mathbf{v}^{(1)}\cdot\bm{\nabla}\theta^{(1)}\rangle-2\lambda\langle\partial_x\theta^{(1)}v_x^{(1)}\rangle-\frac{k_2}{\gamma \, \Er}\langle\partial_y\theta^{(1)}\partial_x\theta^{(1)}\rangle,\label{gg}
\end{eqnarray}
with $k_1=[K_r(1+\lambda)+1-\lambda]/{(2+\mu_2)}$ and $k_2=K_r-1$.
Expanding the no-slip boundary condition to second order, we find
\begin{equation}
\langle v_x ^{(2)} \rangle |_{y=0} = - \langle Y \partial _y v _x ^{(1)} \rangle |_{y=0},
\end{equation}
where $Y$ is given by Eq.~(\ref{yform}).
 The second-order part of the anchoring condition takes the form

\begin{equation}
\left[-\langle\partial_y\theta^{(2)}\rangle+w\langle\theta^{(2)}\rangle\right]_{y=0}=\Xi\label{anchBC2order},
\end{equation}
where 
\begin{equation}
\Xi=\left.\langle -\partial_xY\partial_x\theta^{(1)}+Y\partial_y^2\theta^{(1)}-wY\partial_y\theta^{(1)}\rangle\right|_{y=0}.\label{xverseXi}
\end{equation}
The swimming speed and velocity field at second order are given by solving (\ref{2o1}) and (\ref{2o2}) subject to the no-slip boundary condition and no flow at infinity. The result is
\begin{eqnarray}
\langle v_x ^{(2)} \rangle = \langle v_x ^{(2)}\rangle |_{y=0}- \alpha \int_0 ^{y} \big[ \gamma(1- \lambda)g +  (2  + \mu_2) F\big] dy', \label{intv2eq}
\end{eqnarray}
where $F(y) = \int _y ^{\infty} f(y') dy'$ and $\alpha =  2[\gamma(1-\lambda)^2 +  2(2 + \mu _2)]^{-1}$. The boundary conditions on $\langle\theta^{(2)}\rangle$ do not enter the expression for $\langle v_x^{(2)}\rangle$. The swimming speed $U$ is given by the flow speed (\ref{intv2eq}) at $y=\infty$: 
\begin{eqnarray}
U = \langle v_x ^{(2)} \rangle |_{y=0}- \alpha \int_0 ^{\infty}\big[ \gamma(1-\lambda)g+(2  + \mu_2) y f\big] dy, \label{uintformula}
\end{eqnarray}
\label{Eqn:Ufinal}
where to obtain (\ref{uintformula}) we have integrated by parts. 
Appendix~\ref{detailsappx} discusses some of the details of calculating this integral.

We will also be interested in another observable.  
Unlike in the case of an unconfined Taylor swimmer in a Newtonian~\cite{taylor1951} or Oldroyd-B fluid at zero Reynolds number~\cite{Lauga2007}, there is a net flux of fluid pumped by a swimmer in a liquid crystal. In the lab frame, the average flux is given by 
\begin{eqnarray}
Q=\int_{y_\mathrm{s}}^\infty\langle v_x-U\rangle\mathrm{d}y \approx \int_0^\infty\langle v_x^{(2)}-U\rangle
\mathrm{d}y-\left.\langle y_\mathrm{s}v_x^{(1)}\rangle\right|_{y=0}.
\label{flux}
\end{eqnarray}
Note that the second term of Eqn.~(\ref{flux}) vanishes for a transverse wave since $v_x^{(1)}|_{y=0}=0$. (The second term also vanishes for a longitudinal wave, since $y_\mathrm{s}=0$---see Appendix C.) Therefore, the flux is also given to second-order accuracy by 
\begin{equation}
Q^{(2)}=\int_0^\infty\left(\langle v_x^{(2)}\rangle-U\right)\mathrm{d}y.
\end{equation}
Note our sign convention: a positive $U$ corresponds to swimming towards the left, opposite the direction of wave propagation \ski{(see Fig.~\ref{setup}f)}, while a positive $Q$ corresponds to fluid swept to the right, along the direction of wave propagation.

\begin{figure*}[t]
\includegraphics[width=\textwidth]{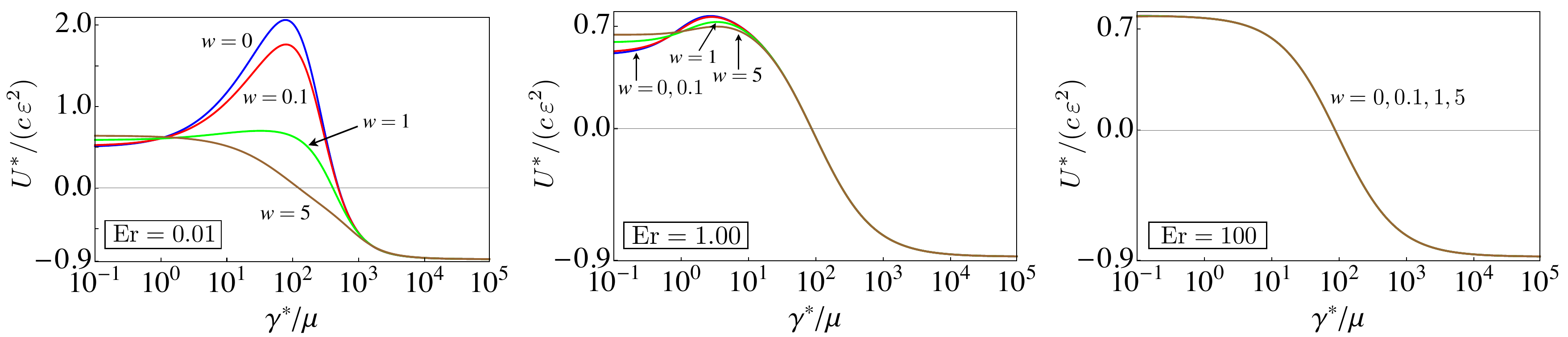}
\caption{(Color online) Dimensionless swimming speed $U^*/(c\varepsilon^2)$  vs. $\gamma^*/\mu$ for a  swimmer in a nematic fluid 
with $\mu_1^* =\mu$, $\mu_2^* = \mu$, $K_r = 1.2$, and $\lambda = 0.75$. The three panels correspond to $\Er=0.01$  (left), $\Er=1.00$ (middle), and $\Er=100.$ (right). The colors denote anchoring strengths: $w=0$ (blue), $w=0.1$ (red), $w=1.0$ (green), and $w=5.0$ (brown). Note that the scale for velocity in the left panel is expanded relative to the scales in the middle and right panels. }
\label{phasegrid}
\end{figure*}

\section{Results}
\label{sec:results}

\subsection{Dependence on Ericksen number}\label{sec:genEr}


For general Ericksen number the solutions of the governing equations to second order do not result in elegant expressions, but the swimming speed and flux are readily found and plotted. {The method of solution is described} in Appendix~\ref{detailsappx}. In the following, we use material parameters that closely mirror the properties of disodium cromolyn glycate (DSCG), in which experiments on swimmers in liquid crystals have been performed \cite{mttwa14,mtwa14,kumar2013LCswim,DSCGprop}. We choose $\mu_1 = \mu_2 = 1$, $K_r= 1.2$, and study the swimming speed as a function of Ericksen number $\Er$, anchoring strength $w$, rotational viscosity $\gamma$, and tumbling parameter $\lambda$.


Figure~\ref{gamtryp} shows the swimming speed 
as a function of $\Er = \tau \omega$ [recall $\tau=\mu/(K_3 q^2)$] and $\gamma^*/\mu=5$, 25, and 50.  The range of $\gamma^*/\mu$ is between $\approx5$ and $\approx50$ for DSCG~\cite{DSCGprop}. Observe that when the anchoring strength is weak, the swimming speed decreases with $\Er$. This behavior was also seen in the case of a swimmer in a hexatic liquid crystal~\cite{ksp14}. When the anchoring strength is strong, $w\gtrapprox5$, the swimming speed is weakly dependent on Ericksen number, becoming independent of Ericksen number for large rotational viscosity (Fig.~\ref{gamtryp}, right panel). This weak dependence on $\Er$ is suggested by the fact that the rotational viscosity enters the governing equations in the combination {$(\gamma\,\Er)^{-1}$} (Eqs.~\ref{thetat1stream}, \ref{2o2}, \ref{gg}). However, when the rotation viscosity is in the low range for DSCG, $\gamma^*/\mu=5$, the swimming speed \textit{increases} with Ericksen number when  the anchoring strength is moderately strong, $w=5$ (Fig.~\ref{gamtryp}, left panel): when anchoring is important, the swimming speed increases when viscous effects dominate as long as the rotational viscosity is sufficiently low. The increase in swimming speed with Er at strong anchoring and modest $\gamma^*$ is not seen in the hexatic liquid crystal~\cite{ksp14}.

All three panels of Fig.~\ref{gamtryp} indicate that the swimming speed becomes independent of anchoring strength when the Ericksen number is sufficiently large. Once again, because the rotational viscosity enters always in the combination {$(\gamma\,\Er)^{-1}$}, the value of $\Er$ for which the anchoring strength becomes irrelevant is inversely proportional to the rotational viscosity. In contrast with the case of a transverse-wave swimmer in a hexatic liquid crystal, the swimming speed has a weak but noticeable dependence on Ericksen number when the anchoring strength is strong and the rotational viscosity is not too {large} (Fig.~\ref{gamtryp}, left panel). But because this dependence is weak we can say that the large Er limit is the same as the strong anchoring strength limit. Note that the large Ericksen number limit is reached at relatively small values of the Ericksen number; in all the panels of Fig.~\ref{gamtryp}, the large Er asymptotic value is reached or nearly reached when $\mathrm{Er}=1$.

As in the case of a hexatic liquid crystal~\cite{ksp14}, the large Er limit is singular, since terms with the highest derivatives in the governing equations vanish in this limit [See Eqs.~(\ref{stokes1stream}--\ref{thetat1stream}),~(\ref{2o1}--\ref{2o2})].  When elastic stresses are small compared with the viscous stresses, it is natural to set the Ericksen number to infinity, or equivalently, drop all terms involving the Frank elastic constants. The resulting limiting model is known as Ericksen's transversely isotropic fluid~\cite{larson1999}. However, this limit is singular, and therefore Ericksen's transversely isotropic fluid does not give physical results for the swimming speed. In particular, Ericksen's transversely isotropic fluid would incorrectly predict that the swimming speed is independent of rotational viscosity at large Ericksen number. Examining the right end of each of the panels in Fig.~\ref{gamtryp} shows that speed depends on $\gamma$ at large Er; we now turn our attention to this dependence.


\begin{figure}[t]
\includegraphics[width=\columnwidth]{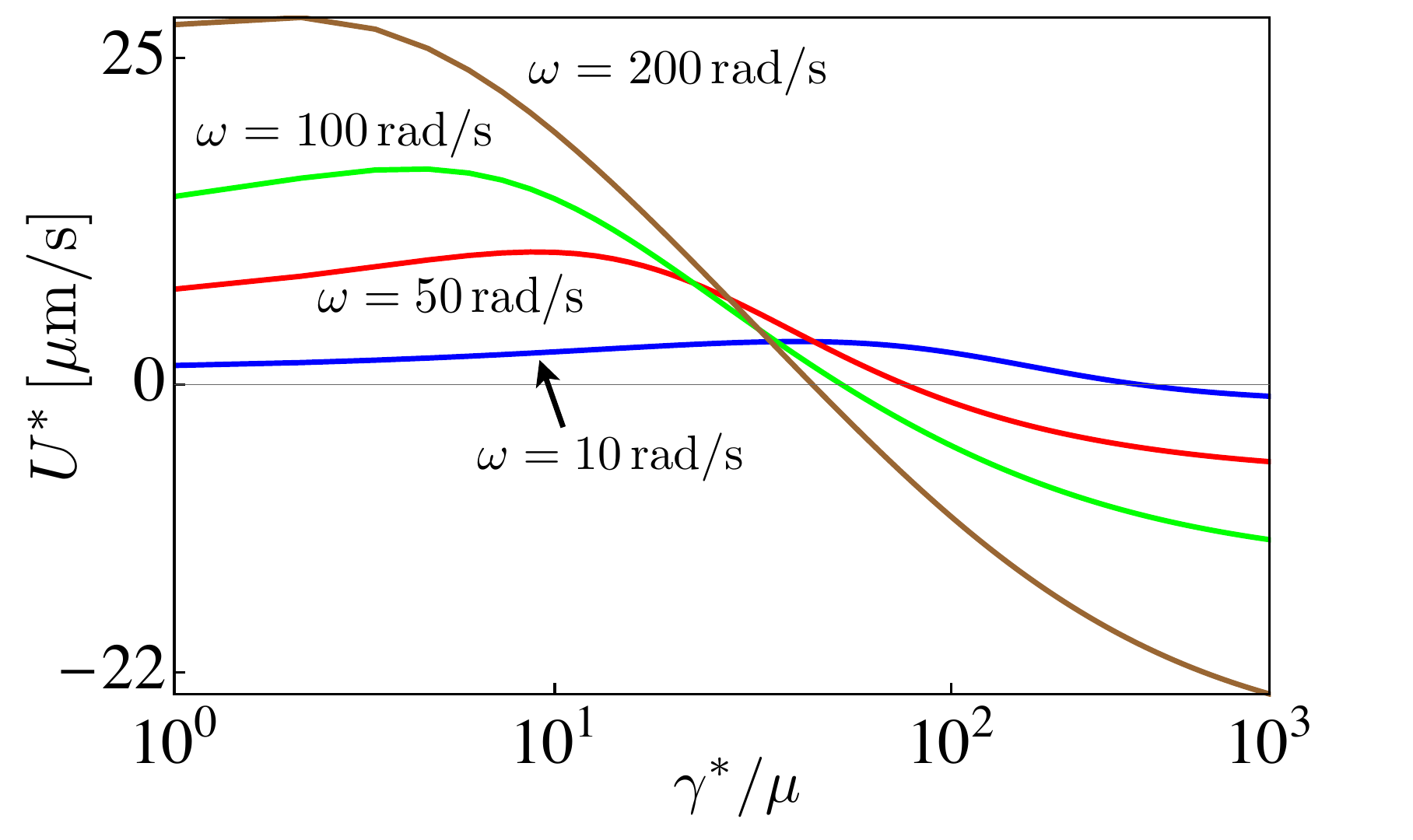}
\caption{(Color online) Swimming speed in $\mu \mathrm{m} \, \mathrm{s}^{-1}$ vs $\gamma^*/\mu$ for a  swimmer in a nematic liquid crystal with $\mu_1 = \mu_2  = 1$, $K_r = 1.2$, $\lambda = 0.75$, $q = 1 \,\mathrm{rad}/\mu \mathrm{m}$, $\tau = 1\,\mathrm{ms}$, and $w=0$. The colored curves correspond to different beat frequencies: $\omega = 10 \, \mathrm{rad/s}$ (blue), $\omega = 50 \, \mathrm{rad/s}$ (red), $\omega = 100 \, \mathrm{rad/s}$ (green), $\omega = 200 \, \mathrm{rad/s}$ (brown). The corresponding values of $\Er = \tau \omega$ are all $\ll 1$.} 
\label{freqdepplot}
\end{figure}

\subsection{Dependence on rotational viscosity $\gamma^*$}
Figure~\ref{phasegrid} shows the swimming speed as a function of dimensionless rotational viscosity $\gamma=\gamma^*/\mu$ for various Ericksen numbers and anchoring strengths for a nematic liquid crystal. 
First note that in accord with discussion of Fig.~\ref{gamtryp}, the swimming speed becomes independent of anchoring strength for large $\Er$. The middle panel of Fig.~\ref{phasegrid} shows that the dependence on anchoring strength is very weak even for $\Er=1$  (which of course can also be observed from Fig.~\ref{gamtryp}). Secondly, the swimmer can reverse direction.  When $\gamma^*/\mu$ is large enough, $U^*$ becomes negative, meaning the swimmer swims in the same direction as its propagating wave. These qualitative features were also observed in the model for a swimmer in a hexatic liquid crystal\cite{ksp14}. However, Fig.~\ref{phasegrid} also reveals important differences between swimming in a hexatic and a nematic liquid crystal. First,  the hexatic swimming speed is always bounded by the isotropic Newtonian swimming speed~\cite{ksp14}, $|U^*|<c\varepsilon^2/2$, whereas the nematic swimming speed can be greater than the Newtonian speed.  Second, there is a maximum in the swimming speed as a function of rotational viscosity, as long as the anchoring strength is low enough. The maximum is most apparent at small Er, and is in the region of measured rotational viscosities for DSCG (Fig~\ref{phasegrid}, left panel). The maximum is less apparent at higher Ericksen numbers since in that regime, the $\gamma\rightarrow0$ limit of the speed is only slightly smaller than the value of the maximum speed. 

Note also that the swimming speed depends on the anchoring strength in the limit of low rotational viscosity. When $\gamma\rightarrow0$, there is a decoupling between the flow field and the director field because the molecular field $\mathbf{h}$ vanishes in this limit. In the problem of swimming in a hexatic liquid crystal~\cite{ksp14}, this decoupling is complete, and the swimming speed is the isotropic Newtonian swimming speed~\cite{taylor1951} $U^*=\omega \varepsilon^2/(2q)$ when $\gamma\rightarrow0$. However, the decoupling is only partial in the case of a nematic liquid crystal, since in that case the anisotropic terms in the viscous stress depend on the director configuration even when $\mathbf{h}=0$. When $\mathbf{h}=0$, the director field at each instant is in equilibrium, and since this equilibrium configuration depends on the anchoring strength, the stress and ultimately the swimming speed depend on the anchoring strength. In particular, the swimming speed does not go to the isotropic swimming speed when $\gamma=0$ in a nematic liquid crystal.

It is interesting to plot the  swimming speed in physical units to make the dependence on beat frequency $\omega$ more  apparent (Fig.~\ref{freqdepplot}). When the anchoring strength vanishes and $\gamma^*/\mu$ is in the experimental range of 10 and 100 for DSCG~\cite{yaoPhD}, the swimming speed depends only weakly on   the beat frequency $\omega$: all four curves in Fig.~\ref{freqdepplot} cross in this region. 

\begin{figure}[t]
\includegraphics[width=\columnwidth]{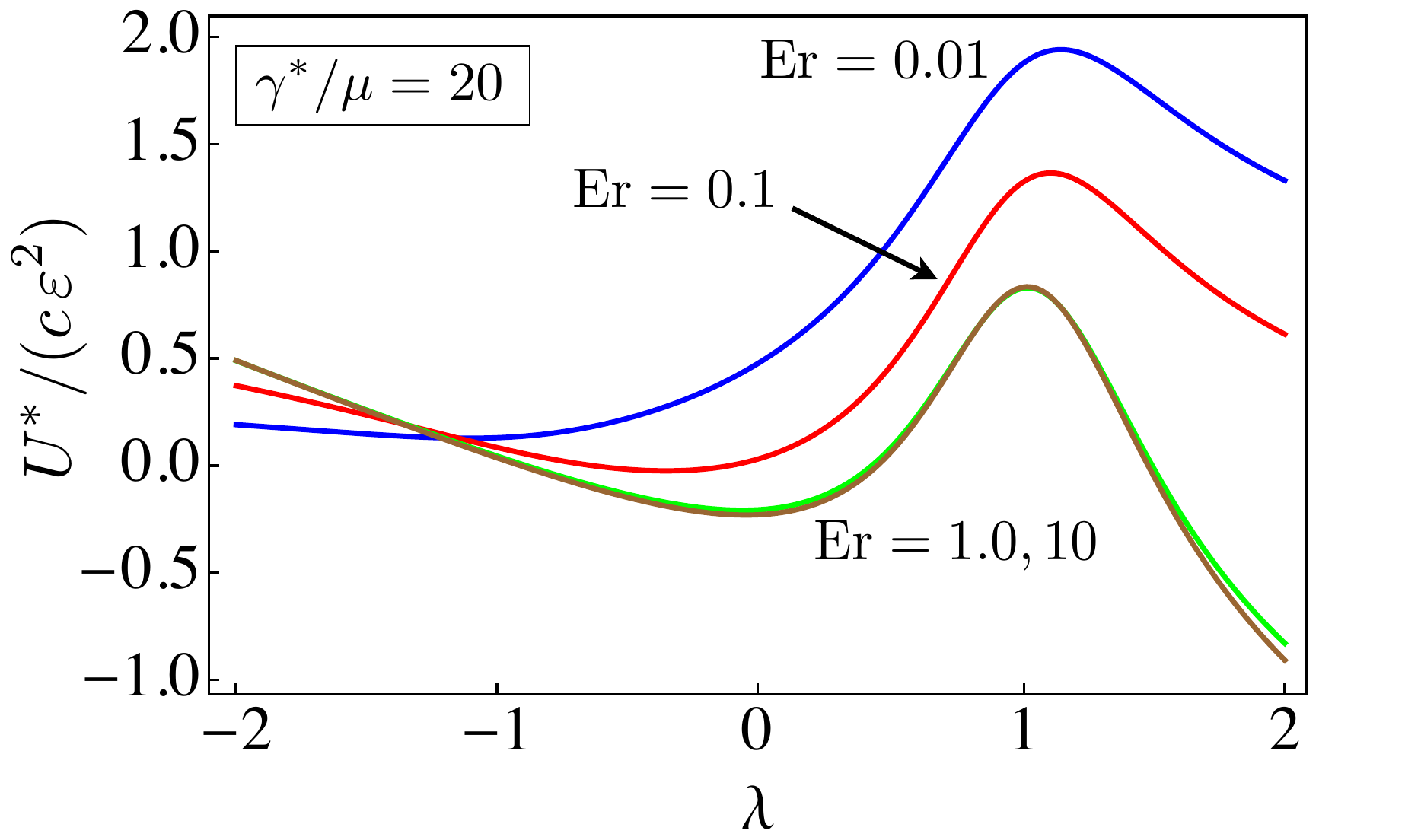}
\caption{(Color online) Dimensionless swimming speed vs tumbling parameter $\lambda$ for various Ericksen numbers,  $\mu_1^*=\mu$, $\mu^*_2=\mu$, $\gamma^*=20\mu$, and zero anchoring strength. }
\label{Uvlam}
\end{figure}
\subsection{Dependence on tumbling parameter $\lambda$}

\begin{figure}[t]
\includegraphics[width=\columnwidth]{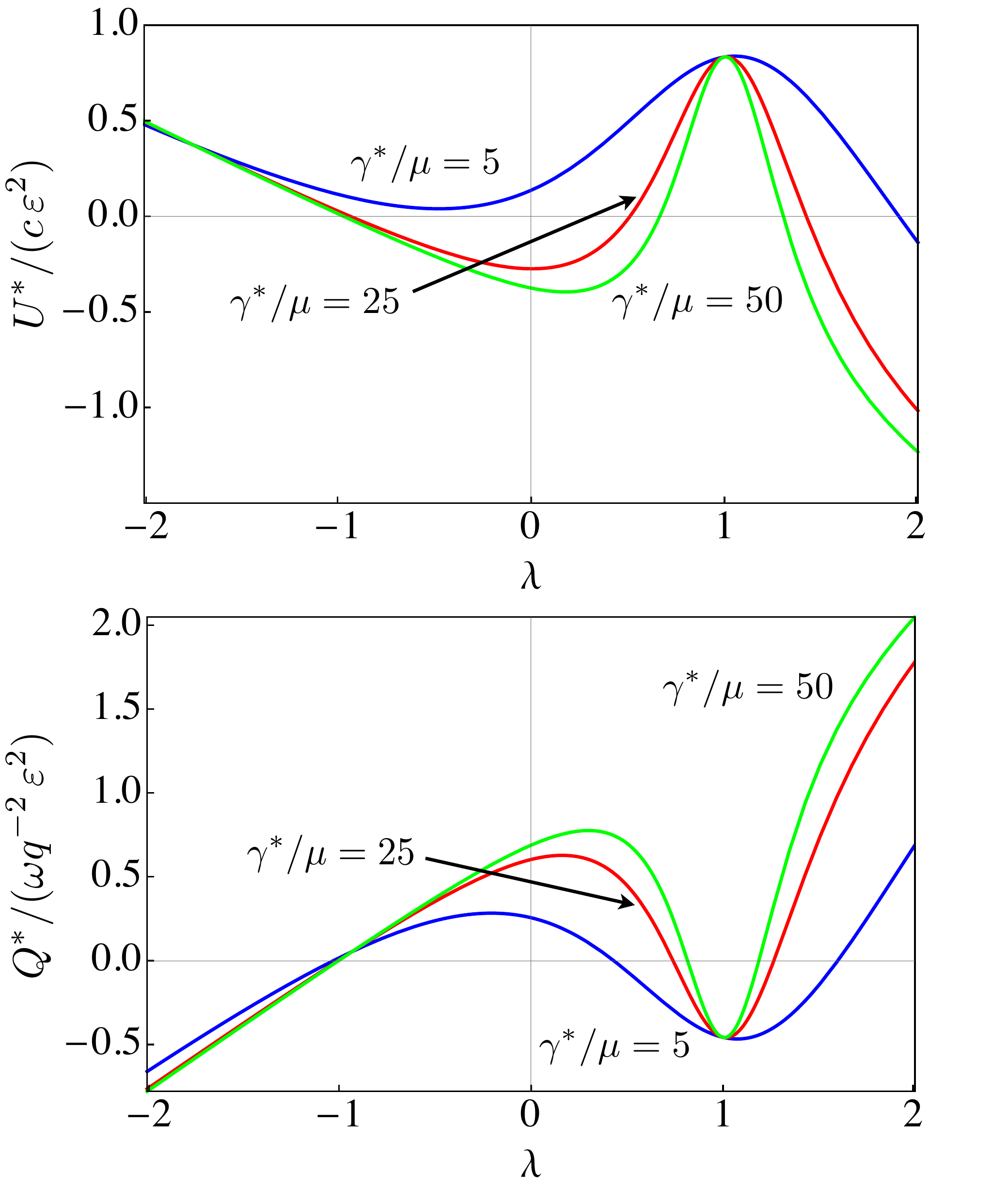}
\caption{(Color online) Dimensionless swimming speed (top) and flux (bottom) vs $\lambda$ at $\Er=1000$  for a  swimmer in a nematic liquid crystal with $\mu_1^*=\mu$, $\mu_2^* = \mu$ and rotational viscosities given by  $\gamma^*/\mu = 5$ (blue), $\gamma^*/\mu = 25$ (red), and $\gamma^*/\mu = 50$ (green).}
\label{ETIFtransqg}
\end{figure}

Figure~\ref{Uvlam} shows how the swimming speed depends on the tumbling parameter $\lambda$ for various Ericksen numbers. For all Ericksen numbers we find a peak near $\lambda=1$, which marks the transition between {tumbling and shear-aligning} nematic liquid crystals~\cite{larson1999}. The maximum is at $\lambda=1$ for moderate to high Er, and shifts to slightly higher $\lambda$ when the Ericksen number becomes small.  
As mentioned earlier, the general expressions for speed and flux are too complicated to display. However, there is a relatively compact expression of the swimming speed in the limit of large Ericksen number, which we find by expanding the first-order solutions of \eqref{rootansatz1}--\eqref{firstanchcond} in a Taylor series in $\mathrm{Er}^{-1}$, and then inserting these values into  \eqref{2o1}--\eqref{uintformula} to find
\begin{equation}
U = \varepsilon^2\frac{2[2+\mu_1(1+\lambda)+\mu_2] -  \gamma (1+\lambda)(\lambda - 1)^2}{8  + 4\mu_2 + 2\gamma(\lambda - 1)^2} + \mathcal{O}\left(\frac{1}{\Er}\right).\label{hiER}
\end{equation}
\color{black}
This expression confirms our general observation that the large-Ericksen number behavior is independent of the anchoring strength $w$. It also shows that when the Ericksen number is large, the swimming speed becomes independent of the rotational viscosity when $\lambda=1$.
The speed and flux as a function of tumbling parameter for various rotational viscosities are plotted in Fig.~\ref{ETIFtransqg}. Note again that although (\ref{hiER}) and Fig.~\ref{ETIFtransqg} are appropriate for large Er, they are applicable even to the modest Ericksen numbers describing experimental systems, of size 1--10, since the swimming speed reaches its high Er limit at a low value of Er. We do not have an explanation for why the swimming speed becomes independent of rotational viscosity when $\lambda=1$, but we offer the following remarks. First, as mentioned previously, the transition from tumbling to {shear}-aligning nematic liquid crystals occurs when $\lambda=1$~\cite{larson1999}. Second, the governing equations  lose some of the highest derivative terms when $\lambda=1$, indicating singular behavior and the existence of boundary layers near the swimmer that are thin in the $y$ direction. And finally, examination of the first order solutions for the angle field in this limit reveal that the angle field simultaneously obeys the strong anchoring and no-anchoring boundary conditions; in other words, the directors align exactly tangential to the swimmer surface, but experience no torque.

\begin{figure}
\includegraphics[width=.95\columnwidth]{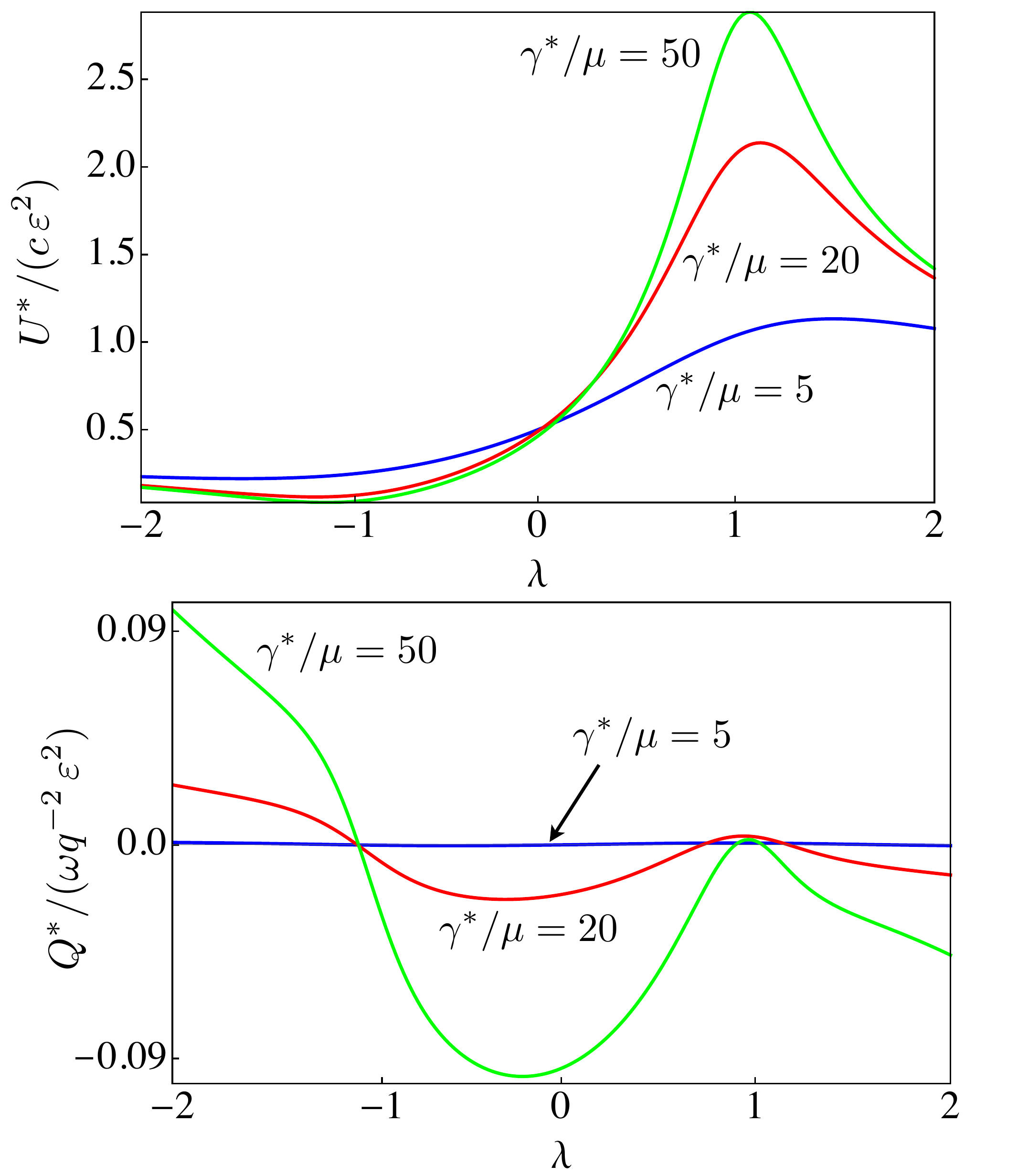}
\caption{(Color online) Dimensionless swimming speed 
vs $\lambda$ for a 
swimmer in a nematic liquid crystal with $\mu_1 = \mu_2  = 1$, $K_r = 1.2$, $w=0$, and $\Er = \tau \omega = 0.01$, with rotational viscosities $\gamma^*/\mu = 5$ (blue), $\gamma^*/\mu = 25$ (red), and $\gamma^*/\mu = 50$ (green).} 
\label{smallerplot}
\end{figure}

We close this section by describing the dependence of speed on tumbling parameter and rotational viscosity for small Ericksen number and weak anchoring strength, which is also an experimentally relevant regime. 
%
%
%
%
%
For the hexatic liquid crystal~\cite{ksp14}, it has been calculated that the first-order velocity field $\mathbf{v}^{(1)}$ is identical to that generated by a swimmer in a Newtonian fluid, so that for $\Er \ll 1$ and $w=0$ the speed is identical to the speed in a Newtonian fluid for any rotational viscosity. In an anisotropic fluid, however, the flow field can differ markedly from the Newtonian counterpart even at first order in $\varepsilon$, which implies that the speed can differ dramatically from the Taylor speed~\cite{taylor1951} $U=c\varepsilon^2/2$, as shown in Fig.~\ref{smallerplot}. In the limit $\lambda \rightarrow 0$, the swimming speed is the same as for a swimmer in an isotropic Newtonian fluid. 
Note however that there is a small but nonzero flux when $\lambda=0$, indicating that the flow field differs from the isotropic flow field. 


\begin{figure}
\includegraphics[width=\columnwidth]{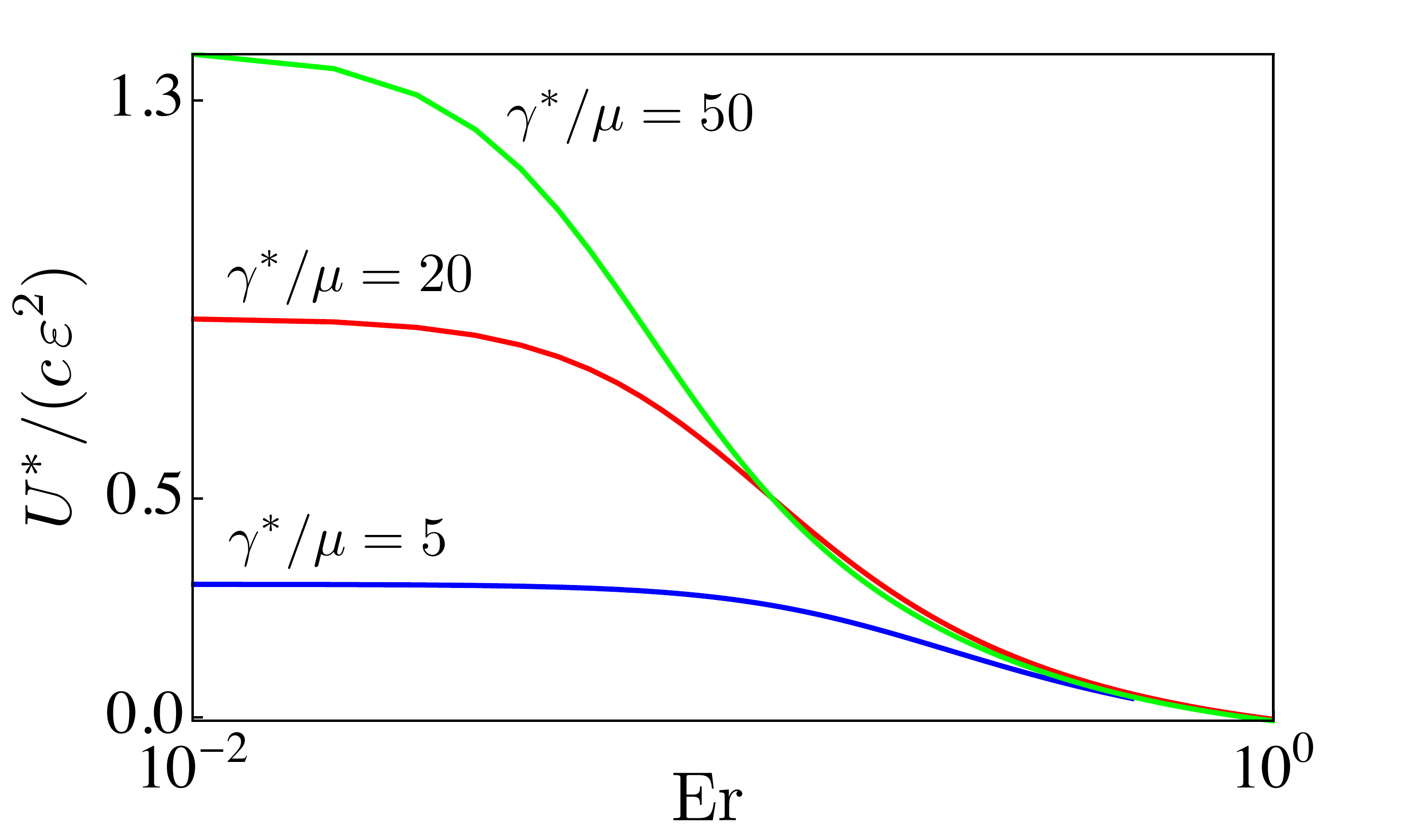}
\caption{(Color online) Dimensionless swimming speed vs Ericksen number $\tau\omega$
for a non-deformable swimmer with prescribed director oscillation and $\mu_1 = \mu_2 = 1$, $K_r = 1.2$, 
and with $\gamma^*/\mu = 5$ (blue), $\gamma^*/\mu= 20$ (red), and $\gamma^*/\mu= 50$ (green).} 
\label{wonk_er}
\end{figure}

\subsection{Swimming and pumping using back flow}\label{sec:noshape}

\TP{To highlight the role of the nematic degree of freedom in our problem, we study swimming and pumping via a mechanism in which all flow is generated by a prescribed motion of the directors at a flat non-deformable wall.
The coupling of the motion of the directors to the flow, and vice-versa, is known as backflow. We suppose that} 
some external mechanism \TP{oscillates the directors along the wall}
with the form of a traveling wave \TP{with wavenumber $q$ and frequency $\omega$}, such that the 
\TP{(dimensionless) boundary conditions at the wall  are
\begin{eqnarray}
\left.\mathbf{v}\right|_{y=0}&=&\mathbf{0}\\
\left.\theta\right|_{y=0}&=&\varepsilon e^{i(x-t)}.
\end{eqnarray}} 
Thus, the director configuration rather than the shape  is prescribed. For brevity, we call the swimmer with prescribed director configuration a `non-deformable' swimmer, and the swimmer with prescribed shape a `deformable' swimmer.

The swimming speed as a function of Ericksen number is shown in Fig.~\ref{wonk_er}.  
A qualitative difference with a deformable swimmer is that the direction does not reverse when the rotational viscosity is large; in fact, increasing the rotational viscosity makes the swimmer go faster, as long as the Ericksen number is not too large. 
Given that the large Ericksen number limit is singular, we expect a boundary layer in the velocity field and angle field when Er is large. In the problem of a deformable swimmer, we found that the swimming speed in that limit is governed by the strong anchoring condition. Since the strong anchoring condition in this case would correspond to no motion of the directors at the swimmer surface, we expect the speed to vanish as Er increases, as our calculations show (Fig.~\ref{wonk_er}). Note also that when $\gamma=0$, there is a complete decoupling between the flow field and director field problems, and the swimming speed vanishes. 

\begin{figure}[t]
\includegraphics[width=\columnwidth]{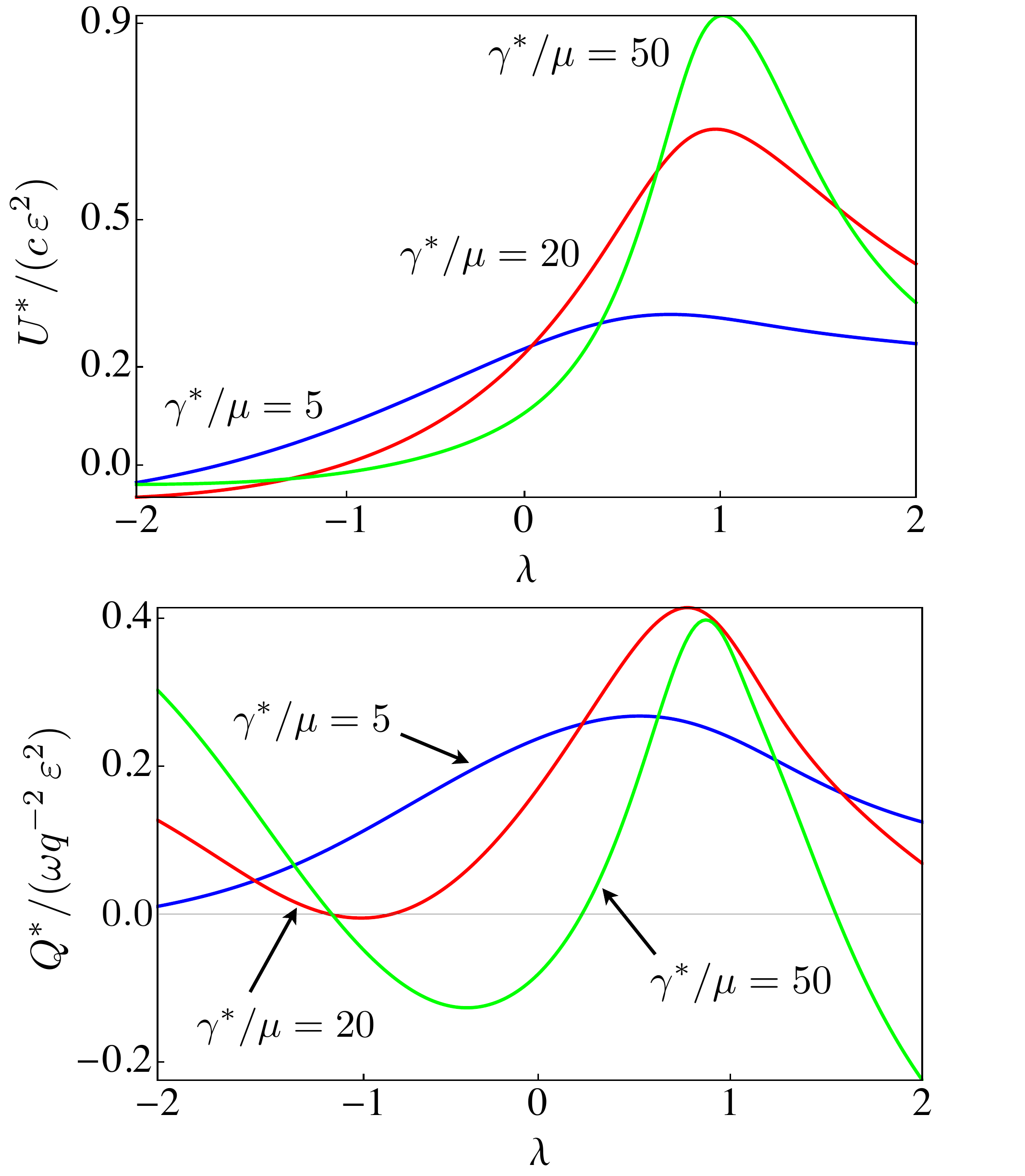}
\caption{(Color online) Dimensionless swimming speed (top) and flux (bottom) in dimensionless units vs $\lambda$ for non-deformable swimmer with prescribed director oscillation
and $\mu_1 = \mu_2 = 1$, $K_r = 1.2$, $\Er = 0.2$ 
and with $\gamma^*/\mu = 5$ (blue), $\gamma^*/\mu= 20$ (red), and $\gamma^*/\mu= 50$ (green).} 
\label{wonkswim}
\end{figure}

\TP{Figure~\ref{wonkswim} shows the $\lambda$-dependence of locomotion and pumping for the non-deformable swimmer. The swimmer can swim faster than the Taylor swimmer when $\lambda\approx1$ and the rotational viscosity is sufficiently large. Note that the behavior of the swimming speed is qualitatively similar to that induced by a swimmer with a deformable body~(Fig.~\ref{smallerplot}).} The flux induced by the motion of the directors in the non-deformable swimmer (Fig~\ref{wonkswim})  is comparable to the flux induced by a deformable swimmer, indicating that at low Ericksen number much of the flux is driven by the backflow effect.

\section{Discussion and Conclusion}
\label{sec:discuss}


Because the nematic phase is more anisotropic than the hexatic phase, more parameters are required in its constitutive relation. In particular, there are anisotropic viscosities as well as different elastic moduli for bend and splay (and twist) 
director configurations. The tumbling parameter $\lambda$ also leads to further distinctions, such as tumbling and {shear}-aligning, which do not exist in the hexatic. Therefore, the hexatic model is good quantitative approximation for swimming in a nematic when the magnitudes of  $\mu_1$, $\mu_2$, $\lambda$, and $K_1/K_3-1$ are small. Except for $K_1/K_3-1$, these parameters are usually not small for the nematic phase.

\begin{figure}[t]
\includegraphics[width=\columnwidth]{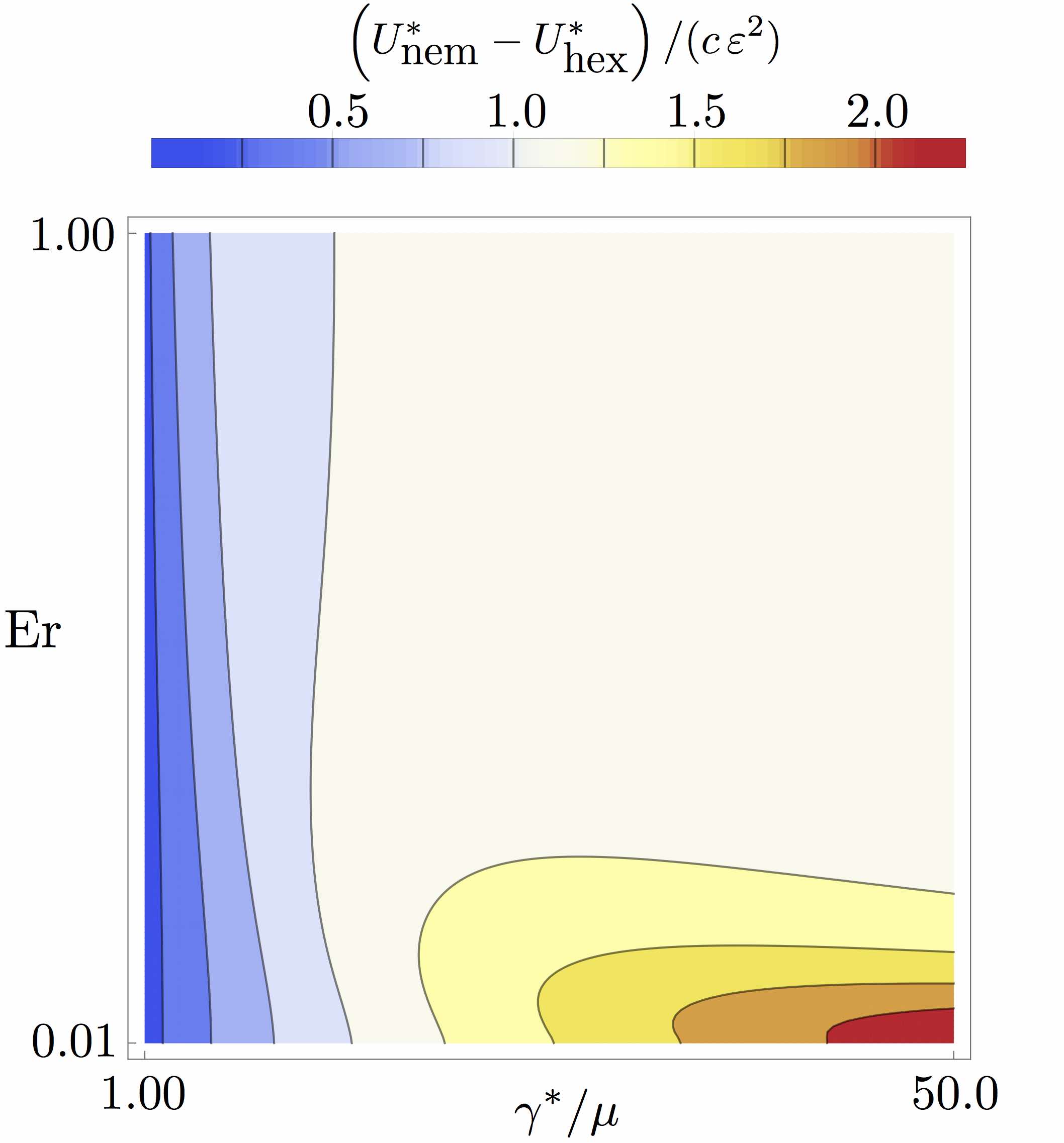}
\caption{(Color online) Difference in non-dimensional swimming speeds versus $\gamma^*/\mu$ and $\Er$ between a highly calamitic nematic fluid ($\lambda = 0.9$) with $\mu_1^*/\mu = \mu_2^*/\mu = 1$, $K_r = 1.2$, and a hexatic fluid. Both fluids are subject to weak anchoring conditions ($w=0$). The horizontal and vertical scales are linear, not logarithmic.} 
\label{compcont}
\end{figure}

\ski{To further highlight the 
quantitative difference in this regime, 
Fig.~\ref{compcont} 
shows the difference in speeds between a highly calamitic, near-aligning transition nematic fluid and a hexatic for small values of the Ericksen number and a range of 
rotational viscosities. For these 
parameters, the swimmer in the nematic fluid can travel at much greater speeds than its companion in a hexatic fluid.}

In this work we extended Taylor's model of an undulating sheet locomoting by means of small-amplitude traveling 
waves in a Newtonian fluid to the case where the ambient fluid is a twist-free nematic liquid crystal.  By considering coupled equations for the local nematic director and velocity fields and expanding perturbatively in 
\TP{the amplitude} we were able to derive general formulas for 
swimming speed 
and volumetric flux induced by the Taylor sheet. 

\TP{Many} 
of the surprising qualitative features, such as reversal of swimming direction for high rotational viscosity, the presence of non-zero volumetrix flux, and a convergence to a strongly-anchored solution for all anchoring strengths at high Ericksen number\TP{, have also been seen in the case of a}
hexatic liquid crystal film~\cite{ksp14}. However, the effects of anisotropy tend for general material parameters to enhance the swimming speed, 
\TP{as can occur in for swimming in porous or elastic fluids~\cite{Leshansky2009,FuShenoyPowers2010}, shear thinning fluids~\cite{RodrigoLauga2013}, or near rigid walls~\cite{Reynolds1965}.}
This speed augmentation by anisotropy can be pronounced, particularly in the low-Ericksen number regime.
\TP{Our results show that the distinctive properties nematic liquid crystals, such as backflow,  can be exploited to develop novel methods of swimming and pumping in anisotropic fluids.}

\section{Acknowledgements}

This work was supported in part by National Science Foundation Grants No. CBET-0854108 (TRP) and CBET-1437195 (TRP). Some of this work was carried out at the Aspen Center for Physics, which is supported by National Science Foundation Grant No. 1066293. We are grateful to John Toner for insightful comments at the early stages of this work, and to Joel Pendery and Marcelo Dias for discussion.

\begin{appendix}

\section{Hexatic equations}
\label{hexaticappx}

For comparison purposes, we include here the governing equations for the hexatic phase:
\begin{eqnarray}
-{\boldsymbol \nabla}p+\mu\nabla^2{\mathbf v}-K{\boldsymbol\nabla}\cdot\left({\boldsymbol \nabla}\theta{\boldsymbol\nabla}\theta\right)+\frac{K}{2}{\boldsymbol\nabla}\times\left(\hat{\mathbf z}\nabla^2\theta\right)={\mathbf 0},\label{hexfbal}\\
\partial_t \theta+\mathbf{v}\cdot{\boldsymbol\nabla}\theta-\frac{1}{2}\hat{\mathbf z}\cdot{\boldsymbol \nabla}\times\mathbf{v}=\frac{K}{\gamma}\nabla^2\theta.\label{tbal}
\end{eqnarray}

%
%

\section{Details of calculating the swimming speed}
\label{detailsappx}

The calculation of the swimming speed, which enters at $O(\e^2)$, depends on a cumbersome but straight-forward combination of the first-order flow and director fields via \eqref{Eqn:Ufinal}. The real part of the first order stream-function in \eqref{rootansatz1} may be written as
\begin{gather}
\mathfrak{R}[\tilde{\psi}^{(1)}] = \frac{1}{2} \left(\tilde{\psi}^{(1)}+\overline{\tilde{\psi}^{(1)}}\right),
\end{gather}
where the overbar denotes the complex conjugate. The director angle field is similarly treated. In \eqref{ff}-\eqref{gg} we require such quantities as $\langle \theta_x\theta_{yy} \rangle$, which we obtain via
\begin{multline}
\theta_x\theta_{yy}  = \sum_{j,k=1}^3 \left(i d_j e^{r_j y   +  \color{black} i(x-t)}-i \bar{d_j} e^{\bar{r_j} y   -  \color{black} i(x-t)}\right) \times\\
\left(d_k r_k^2 e^{r_k y   +  \color{black} i(x-t)}+\bar{d_k} \bar{r_k}^2 e^{\bar{r_k} y   -  \color{black} i(x-t)}\right).
\end{multline}
The horizontal mean over one period is then given by
\begin{gather}
I_1=\langle \theta_x\theta_{yy} \rangle= \frac{i}{4}\sum_{j,k=1}^3\left(d_j \bar{d_k}\bar{r_k}^2 e^{(r_j+\bar{r_k})y} -\bar{d_j} d_k r_k^2  e^{(\bar{r_j} +r_k) y}\right).
\end{gather}
The final integration in \eqref{uintformula} is now easily performed; for example, we have
\begin{gather}
\int_0^\infty y\, I_1\,dy =\frac{i}{4} \sum_{j,k=1}^3 \left[\frac{d_j \bar{d_k}\bar{r_k}^2}{(r_j +\bar{r_k})^2} -\frac{\bar{d_j} d_k r_k^2}{(\bar{r_j} +r_k)^2} \right],
\end{gather}
and the other contributions are deduced in the same fashion. The end result is a cumbersome algebraic expression but one that is easily evaluated for all parameter values.  

\begin{figure}
\includegraphics[width=\columnwidth]{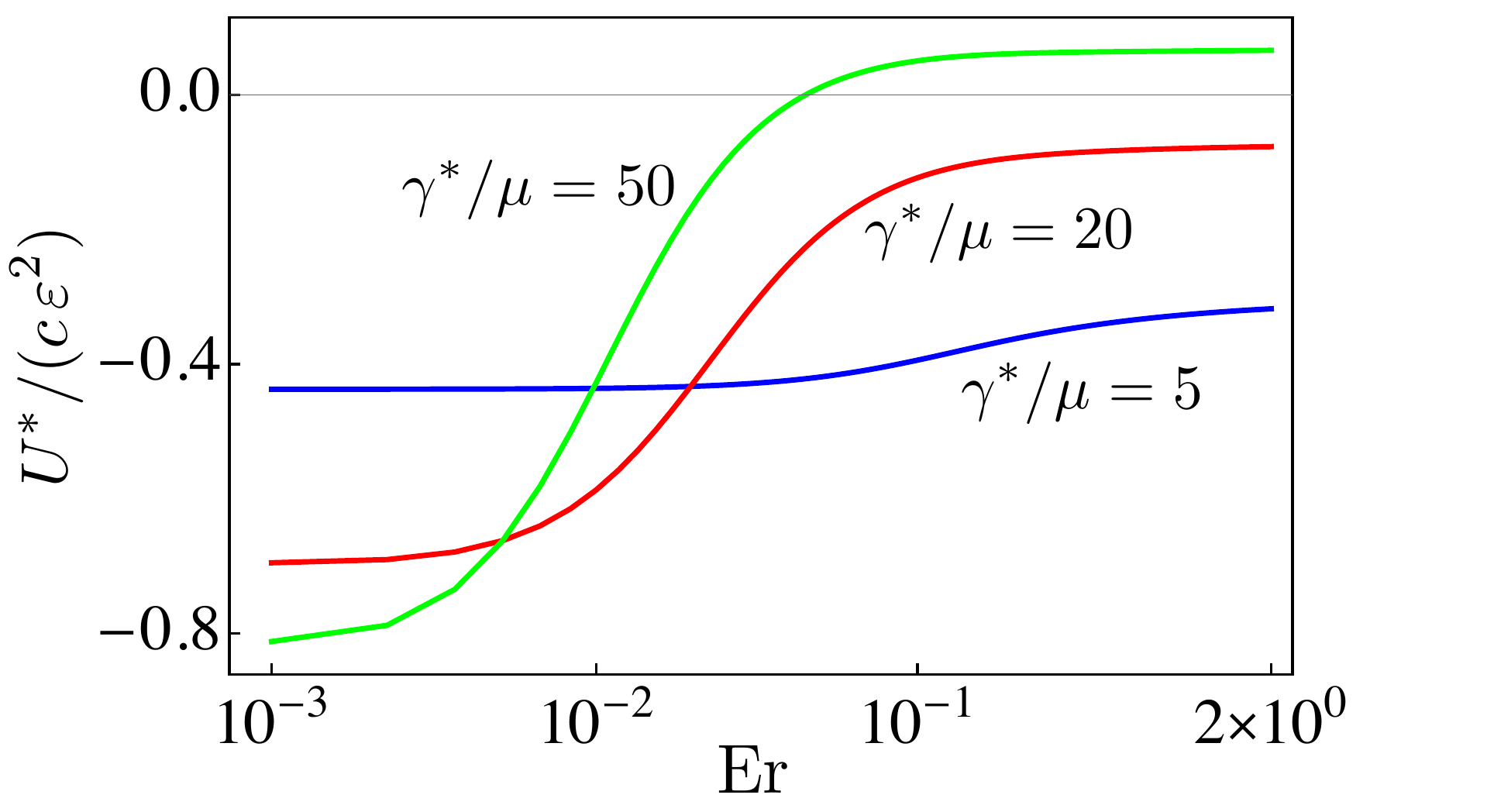}
\caption{(Color online) Dimensionless swimming speed vs. Er for various rotational viscosities for a swimmer with a longitudinal wave in a nematic liquid crystal with $ \mu_1^*=\mu$ , $\mu_2^* = \mu$, $K_r = 1.2$, $\lambda = 0.6$, and zero anchoring strength $w$.} \label{longplot}
\end{figure}

\section{Swimmer with a longitudinal wave}

For completeness, we also present some results for a swimmer with a longitudinal waveform, 
\begin{equation}
(X,Y) = (\varepsilon, 0)\sin(x- t),\label{xform}
\end{equation}
in dimensionless form. Many of the equations needed for calculating the speed and flux are the same as in the case of the transverse swimmer. Here we list the equations that must be modified. The boundary conditions at first order in amplitude $\e$ for the longitudinal swimmer are
\begin{eqnarray}
\partial _y \tilde{\psi} ^{(1)} |_{y=0} &=& - \varepsilon e^{i(x-t)}, \\
-\partial _x \tilde{\psi}^{(1)} |_{y=0} &= &0, \\
 -\partial _y \tilde{\theta}^{(1)} + w \tilde{\theta}^{(1)} |_{y=0}& =& 0\label{firstanchcondl}.
\end{eqnarray}
The boundary condition for the flow field at second order is
\begin{equation}
\langle v_x ^{(2)} \rangle |_{y=0} = - \langle X \partial _y v _x ^{(1)} \rangle |_{y=0},
\end{equation}
with $X$  given by \eqref{xform}.
The anchoring condition remains the same as Eq.~(\ref{anchBC2order}),  with the replacement of $\Xi$  [Eq.~(\ref{xverseXi})] with 
\begin{equation}
\Xi=\left.\langle  X\partial_x\partial_y\theta^{(1)}-w X\partial_x\theta^{(1)}\rangle\right|_{y=0}.
\end{equation}

The swimming  speed vs Ericksen number is shown in Fig.~\ref{longplot}. For most values of Er and rotational viscosity, the swimming speed is negative, meaning the swimmer moves in the same direction as the propagating longitudinal wave, just as in the case of a longitudinal swimmer in an isotropic Newtonian liquid, where the swimming speed is $U^*=-c\varepsilon^2/2$. As in the transverse case, the swimming direction can reverse if the rotational viscosity is sufficiently high. There is no simple formula for the swimming speed for generic values of the parameters, but the swimming velocity at high Ericksen number takes a simple form, with the speed exactly as the transverse case but with opposite direction:
\begin{equation}
U = -\varepsilon^2\frac{2[2+\mu_1(1+\lambda)+\mu_2] -  \gamma (1+\lambda)(\lambda - 1)^2}{8  + 4\mu_2 + 2\gamma(\lambda - 1)^2} + \mathcal{O}\left(\frac{1}{\Er}\right).
\end{equation}
\color{black}
 The entrained flux in this limit is likewise of same magnitude but opposite direction

The longitudinal swimmer in a nematic is slower than the transverse swimmer. 
However, a longitudinal swimmer in a nematic is very different from a longitudinal swimmer in a hexatic. In the case of swimming in a hexatic liquid crystal, the longitudinal swimmer's speed differs from the isotropic speed by only a few percent~\cite{ksp14}. Figure~\ref{longplot} shows that the difference between the isotropic and nematic swimming speed can be significant, especially at higher values of rotational viscosity.

\end{appendix}


\footnotesize{
\bibliography{newrefs}
\bibliographystyle{rsc} 
}

\end{document}